\documentclass{spie}%
\usepackage{amsfonts}
\usepackage{amsmath}
\usepackage{amssymb}
\usepackage{graphicx}%
\newtheorem{proposition}[theorem]{Proposition}
\begin{document}

\title{Continuous Quantum Hidden Subgroup Algorithms}
\author{Samuel J. Lomonaco, Jr.\supit{a} and Louis H. Kauffman\supit{b}
\skiplinehalf\supit{a}University of Maryland Baltimore County, Baltimore,
Maryland 21250, USA\\\supit{b}University of Illinois at Chicago,Chicago, Illinois 60607-7045, USA}
\authorinfo{Further author information: S.J.L.E-mail: lomonaco@umbc.edu, L.H.K. E-mail:
kauffman@uic.edu; S.J.L. URL: www.csee.umbc.edu/\symbol{126}lomonaco, L.H.K.
URL: www.math.uic.edu/\symbol{126}kauffman}
\maketitle

\begin{abstract}
In this paper we show how to construct two continuous variable and one
continuous functional quantum hidden subgroup (QHS) algorithms. \ These are
respectively quantum algorithms on the additive group of reals $\mathbb{R}$,
the additive group $\mathbb{R}/\mathbb{Z}$ of the reals $\mathbb{R}%
\operatorname{mod}1$, i.e., the circle, and the additive group $\textsc{Paths}%
$ of $L^{2}$ paths $x:\left[  0,1\right]  \rightarrow\mathbb{R}^{n}$ in real
$n$-space $\mathbb{R}^{n}$. \ Also included is a curious discrete QHS
algorithm which is dual to Shor's algorithm.

\end{abstract}
\tableofcontents

\section{Introduction}

In this paper we show how to construct two continuous variable and one
continuous functional quantum hidden subgroup (QHS) algorithms. \ These are
quantum algorithms \textsc{Alg}$_{\mathbb{R}}$, \textsc{Alg}$_{\mathbb{R}%
/\mathbb{Z}}$, and \textsc{Alg}$_{\text{\textsc{Paths}}}$, respectively on the
additive group of reals $\mathbb{R}$, the additive group $\mathbb{R}%
/\mathbb{Z}$ of the reals $\mathbb{R}\operatorname{mod}1$, i.e., the circle,
and the additive group \textsc{Paths} of $L^{2}$ paths $x:\left[  0,1\right]
\rightarrow\mathbb{R}^{n}$ in real $n$-space $\mathbb{R}^{n}$. \ With the
methods found in a recent paper\cite{Lomonaco1}, it is a straight forward
exercise to extend \textsc{Alg}$_{\mathbb{R}}$ and \textsc{Alg}$_{\mathbb{R}%
/\mathbb{Z}}$ to a wandering algorithms respectively on the additive group of
real n-tuples $\mathbb{R}^{n}$, and the additive group of the $n$-dimensional
torus $T^{n}=\oplus_{1}^{n}\mathbb{R}/\mathbb{Z}$. \ 

The chief advantage of the QHS algorithm \textsc{Alg}$_{\mathbb{R}/\mathbb{Z}%
}$ over \textsc{Alg}$_{\mathbb{R}}$ is that the ambient space $\mathbb{R}%
/\mathbb{Z}$\ is compact. \ As a result, \textsc{Alg}$_{\mathbb{R}/\mathbb{Z}}
$ can easily be approximated by a sequence \textsc{Alg}$_{\mathbb{Z}_{Q}}$ of
QHS algorithms over suitably chosen finite groups.\ (Each of these algorithms
\textsc{Alg}$_{\mathbb{Z}_{Q}}$ is a natural dual to Shor's algorithm.) The
last QHS algorithm \textsc{Alg}$_{\text{\textsc{Paths}}}$ is a quantum
functional integral algorithm which is highly speculative. The algorithm, in
the spirit of Feynman, is based on functional integrals whose existence is
difficult to determine, let alone approximate. \ However, in the light of
Witten's functional integral approach\cite{Witten1} to the knot invariants,
this algorithm has the advantage of suggesting a possible approach toward
developing a QHS algorithm for the the Jones polynomial.

\bigskip

\section{Definition of the hidden subgroup problem (HSP) and hidden subgroup
algorithms (HSAs)}

\qquad\bigskip

We now proceed by defining what is meant by a hidden subgroup problem (HSP)
and a corresponding hidden subgroup algorithm. \ There are other perspectives
to be found in the open literature.\cite{Kitaev1}\ \ \cite{Jozsa4}%
\ \ \cite{Mosca1}

\bigskip

\begin{definition}
A map $\varphi:A\longrightarrow S$ from a group $A$ into a set $S$ is said to
have \textbf{hidden subgroup structure} if there exists a subgroup
$K_{\varphi}$ of $A$, called a \textbf{hidden subgroup}, and an injection
$\iota_{\varphi}:A/K_{\varphi}\longrightarrow S$, called a \textbf{hidden
injection}, such that the diagram
\[%
\begin{array}
[c]{ccc}%
A\  & \overset{\varphi}{\longrightarrow} & S\\
\nu\searrow &  & \nearrow\iota_{\varphi}\\
& A/K_{\varphi} &
\end{array}
\]
is commutative, where $A/K_{\varphi}$ denotes the collection of right cosets
of $K_{\varphi}$ in $A$, and where $\nu:A\longrightarrow A/K_{\varphi}$ is the
natural map of $A$ onto $A/K_{\varphi}$. \ We refer to the group $A$ as the
\textbf{ambient group} and to the set $S$ as the \textbf{target set}. \ If
$K_{\varphi}$ is a normal subgroup of $A$, then $H_{\varphi}=A/K_{\varphi} $
is a group, called the \textbf{hidden quotient group}, and $\nu
:A\longrightarrow A/K_{\varphi}$ is an epimorphism, called the \textbf{hidden
epimorphism}.
\end{definition}

\bigskip

The hidden subgroup problem can be expressed as follows:

\bigskip

\noindent\textbf{Hidden Subgroup Problem (HSP).} \ \textit{Given a map with
hidden subgroup structure }%
\[
\varphi:A\longrightarrow S\text{ ,}%
\]
\textit{determine a hidden subgroup }$K_{\varphi}$\textit{\ of }$A$\textit{.
\ \ An algorithm solving this problem is called a \textbf{hidden subgroup
algorithm (HSA)}.}

\bigskip

The corresponding quantum form of this HSP is:

\bigskip

\noindent\textbf{Hidden Subgroup Problem:\ Quantum Version.} \ \textit{Let }%
\[
\varphi:A\longrightarrow S
\]
\textit{be a map with hidden subgroup structure. \ Construct a quantum
implementation of the map }$\varphi$\textit{\ as follows: \ \medskip}

\textit{Let }$\mathcal{H}_{A}$\textit{\ and }$\mathcal{H}_{S}$\textit{\ be
Hilbert spaces (or rigged Hilbert spaces) defined respectively by the
orthonormal bases }%
\[
\left\{  \ \left\vert a\right\rangle \mid a\in A\ \right\}  \text{ and
}\left\{  \ \left\vert s\right\rangle \mid s\in S\ \right\}  \text{ ,}%
\]
\textit{and let }$s_{0}=\varphi\left(  0\right)  $\textit{, where }%
$0$\textit{\ denotes the identity of the ambient group }$A$\textit{.
\ Finally, let }$U_{\varphi}$\textit{\ be a unitary transformation such that }%
\[%
\begin{array}
[c]{ccc}%
U_{\varphi}:\mathcal{H}_{A}\otimes\mathcal{H}_{S} & \longrightarrow &
\mathcal{H}_{A}\otimes\mathcal{H}_{S}\\
&  & \\
\left\vert a\right\rangle \left\vert s_{0}\right\rangle  & \longmapsto &
\left\vert a\right\rangle \left\vert \varphi\left(  a\right)  \right\rangle
\end{array}
,
\]
\bigskip

\section{A QHS Algorithm \textsc{Alg}$_{\mathbb{R}}$ on the reals $\mathbb{R}
$}

\bigskip

Let
\[
\varphi:\mathbb{R}\longrightarrow\mathbb{C}%
\]
be a periodic admissible function of minimum period $P$ from the reals
$\mathbb{R}$ to the complex numbers $\mathbb{C}$. \ We will now create a
continuous variable Shor algorithm to find integer periods. \ This algorithm
can be extended to rational and irrational periods \cite{Lomonaco4}. \ 

\bigskip

We construct two quantum registers
\[
\left\vert \text{\textsc{Left Register}}\right\rangle \text{ and }\left\vert
\text{\textsc{Right Register}}\right\rangle
\]
called left- and right-registers respectively, and `living' respectively in
the rigged Hilbert spaces $\mathcal{H}_{\mathbb{R}}$ and $\mathcal{H}%
_{\mathbb{C}}$. \ The left register was constructed to hold arguments of the
function $\varphi$, the right to hold the corresponding function values.

\bigskip

We assume we are given the unitary transformation
\[
U_{\varphi}:\mathcal{H}_{\mathbb{R}}\otimes\mathcal{H}_{\mathbb{C}%
}\longrightarrow\mathcal{H}_{\mathbb{R}}\otimes\mathcal{H}_{\mathbb{C}}%
\]
defined by
\[
U_{\varphi}:\left\vert x\right\rangle \left\vert y\right\rangle \longmapsto
\left\vert x\right\rangle \left\vert y+\varphi\left(  x\right)  \right\rangle
\]

\bigskip

Finally, we choose a large positive integer $Q$, so large that $Q\geq2P^{2}$.

\bigskip

The quantum part of our algorithm consists of \fbox{\textbf{Step 0}} through
\fbox{\textbf{Step 4}} as described below:

\bigskip

\begin{itemize}
\item[\fbox{\textbf{Step 0}}] Initialize
\[
\left\vert \psi_{0}\right\rangle =\left\vert 0\right\rangle \left\vert
0\right\rangle
\]

\item[\fbox{\textbf{Step 1}}] Apply the inverse Fourier transform to the left
register, i.e. apply $\mathcal{F}^{-1}\otimes1$ to obtain
\[
\left\vert \psi_{1}\right\rangle =%
%TCIMACRO{\dint \limits_{-\infty}^{\infty}}%
%BeginExpansion
{\displaystyle\int\limits_{-\infty}^{\infty}}
%EndExpansion
dx\ e^{2\pi ix\cdot0}\left\vert x\right\rangle \left\vert 0\right\rangle =%
%TCIMACRO{\dint \limits_{-\infty}^{\infty}}%
%BeginExpansion
{\displaystyle\int\limits_{-\infty}^{\infty}}
%EndExpansion
dx\ \left\vert x\right\rangle \left\vert 0\right\rangle
\]

\item[\fbox{\textbf{Step 2}}] Apply $U_{\varphi}:\left\vert x\right\rangle
\left\vert u\right\rangle \longmapsto\left\vert x\right\rangle \left\vert
u+\varphi\left(  x\right)  \right\rangle $ to obtain
\[
\left\vert \psi_{2}\right\rangle =%
%TCIMACRO{\dint \limits_{-\infty}^{\infty}}%
%BeginExpansion
{\displaystyle\int\limits_{-\infty}^{\infty}}
%EndExpansion
dx\ \left\vert x\right\rangle \left\vert \varphi\left(  x\right)
\right\rangle
\]

\item[\fbox{\textbf{Step 3}}] Apply the Fourier transform to the left
register, i.e. apply $\mathcal{F}\otimes1$ to obtain
\begin{align*}
\left\vert \psi_{3}\right\rangle  &  =%
%TCIMACRO{\dint \limits_{-\infty}^{\infty}}%
%BeginExpansion
{\displaystyle\int\limits_{-\infty}^{\infty}}
%EndExpansion
dy%
%TCIMACRO{\dint \limits_{-\infty}^{\infty}}%
%BeginExpansion
{\displaystyle\int\limits_{-\infty}^{\infty}}
%EndExpansion
dx\ e^{-2\pi ixy}\left\vert y\right\rangle \left\vert \varphi\left(  x\right)
\right\rangle =%
%TCIMACRO{\dint \limits_{-\infty}^{\infty}}%
%BeginExpansion
{\displaystyle\int\limits_{-\infty}^{\infty}}
%EndExpansion
dy\ \left\vert y\right\rangle \delta_{P}\left(  y\right)
%TCIMACRO{\dint \limits_{0}^{P-}}%
%BeginExpansion
{\displaystyle\int\limits_{0}^{P-}}
%EndExpansion
dx\ e^{-2\pi ixy}\left\vert \varphi\left(  x\right)  \right\rangle \\
& \\
&  =%
%TCIMACRO{\dsum \limits_{n=-\infty}^{\infty}}%
%BeginExpansion
{\displaystyle\sum\limits_{n=-\infty}^{\infty}}
%EndExpansion
\left\vert \frac{n}{P}\right\rangle \left(  \frac{1}{\left\vert P\right\vert }%
%TCIMACRO{\dint \limits_{0}^{P-}}%
%BeginExpansion
{\displaystyle\int\limits_{0}^{P-}}
%EndExpansion
dx\ e^{-2\pi ix\frac{n}{P}}\left\vert \varphi\left(  x\right)  \right\rangle
\right)  =%
%TCIMACRO{\dsum \limits_{n=-\infty}^{\infty}}%
%BeginExpansion
{\displaystyle\sum\limits_{n=-\infty}^{\infty}}
%EndExpansion
\left\vert \frac{n}{P}\right\rangle \left\vert \Omega\left(  \frac{n}%
{P}\right)  \right\rangle
\end{align*}
where
\[
\left\vert \Omega\left(  \frac{n}{P}\right)  \right\rangle =\frac
{1}{\left\vert P\right\vert }%
%TCIMACRO{\dint \limits_{0}^{P-}}%
%BeginExpansion
{\displaystyle\int\limits_{0}^{P-}}
%EndExpansion
dx\ e^{-2\pi ix\frac{n}{P}}\left\vert \varphi\left(  x\right)  \right\rangle
\text{ ,}%
\]
and where $%
%TCIMACRO{\dint \limits_{0}^{P-}}%
%BeginExpansion
{\displaystyle\int\limits_{0}^{P-}}
%EndExpansion
dx\ e^{-2\pi ix\frac{n}{P}}\left\vert \varphi\left(  x\right)  \right\rangle $
denotes the formal integral $%
%TCIMACRO{\dint \limits_{0}^{P}}%
%BeginExpansion
{\displaystyle\int\limits_{0}^{P}}
%EndExpansion
dx\ e^{-2\pi ix\frac{n}{P}}\left[  1-\delta\left(  x-P\right)  \right]
\left\vert \varphi\left(  x\right)  \right\rangle $.
\end{itemize}

\bigskip

\begin{itemize}
\item[\fbox{\textbf{Step 4}}] Measure the left register with respect to the
observable\footnote{$\left\lfloor Qy\right\rfloor $ denotes the greatest
integer $\leq Qy$.}
\[
\mathcal{O}=%
%TCIMACRO{\dint \limits_{-\infty}^{\infty}}%
%BeginExpansion
{\displaystyle\int\limits_{-\infty}^{\infty}}
%EndExpansion
dy\ \frac{\left\lfloor Qy\right\rfloor }{Q}\left\vert y\right\rangle
\left\langle y\right\vert
\]
to produce a random eigenvalue
\[
\frac{m}{Q}\text{ }%
\]
for which there exists a rational of the form $\frac{n}{P}$ such that
$\frac{m}{Q}\leq\frac{n}{P}<\frac{m+1}{Q}$. \ 

\item[\fbox{\textbf{Step 5}}] If $Q\geq2P^{2}$, then $\frac{n}{P}$ is unique
and is a convergent of the continued fraction expansion of $\frac{m}{Q}$.
\ Thus $\frac{n}{P}$ can be computed from the standard continued fraction recursion.
\end{itemize}

\bigskip

\section{A QHS algorithm \textsc{Alg}$_{\mathbb{R}/\mathbb{Z}}$ on the circle
$\mathbb{R}/\mathbb{Z}$}

\bigskip

Given a periodic admissible function with rational period $P$ from the circle
$\mathbb{R}/\mathbb{Z}$ to the complex plane $\mathbb{C}$%
\[
\varphi:\mathbb{R}/\mathbb{Z}\longrightarrow\mathbb{C}\text{,}%
\]
\ we will now create a QHS algorithm to find its period . \ This algorithm can
be extended to irrational periods\cite{Lomonaco5}.

\bigskip

\begin{proposition}
If $\varphi:\mathbb{R}/\mathbb{Z}\longrightarrow\mathbb{C}$ has a rational
period, then its minimum non-trivial period $P$ is a reciprocal integer
$P=\frac{1}{a}\operatorname{mod}1$.\ 
\end{proposition}

\bigskip

We assume we are given the unitary transformation
\[
U_{\varphi}:\mathcal{H}_{\mathbb{R}/\mathbb{Z}}\otimes\mathcal{H}_{\mathbb{C}%
}\longrightarrow\mathcal{H}_{\mathbb{R}/\mathbb{Z}}\otimes\mathcal{H}%
_{\mathbb{C}}%
\]
defined by
\[
U_{\varphi}:\left\vert x\right\rangle \left\vert y\right\rangle \longmapsto
\left\vert x\right\rangle \left\vert y+\varphi\left(  x\right)
\operatorname{mod}1\right\rangle
\]

\bigskip

The quantum part of our algorithm consists of \fbox{\textbf{Step 0}} through
\fbox{\textbf{Step 4}} as described below:

\bigskip

\bigskip

\begin{itemize}
\item[\fbox{\textbf{Step 0}}] Initialize
\[
\left\vert \psi_{0}\right\rangle =\left\vert 0\right\rangle \left\vert
0\right\rangle \in\mathcal{H}_{\mathbb{Z}}\otimes\mathcal{H}_{\mathbb{C}}%
\]

\item[\fbox{\textbf{Step 1}}] Apply the inverse Fourier transform to the left
register, i.e. apply $\mathcal{F}^{-1}\otimes1$ to obtain
\[
\left\vert \psi_{1}\right\rangle =%
%TCIMACRO{\doint }%
%BeginExpansion
{\displaystyle\oint}
%EndExpansion
dx\ e^{2\pi ix\cdot0}\left\vert x\right\rangle \left\vert 0\right\rangle =%
%TCIMACRO{\doint }%
%BeginExpansion
{\displaystyle\oint}
%EndExpansion
dx\ \left\vert x\right\rangle \left\vert 0\right\rangle \in\mathcal{H}%
_{\mathbb{R}/\mathbb{Z}}\otimes\mathcal{H}_{\mathbb{C}}%
\]

\item[\fbox{\textbf{Step 2}}] Apply $U_{\varphi}:\left\vert x\right\rangle
\left\vert u\right\rangle \longmapsto\left\vert x\right\rangle \left\vert
u+\varphi\left(  x\right)  \operatorname{mod}1\right\rangle $ to obtain
\[
\left\vert \psi_{2}\right\rangle =%
%TCIMACRO{\doint }%
%BeginExpansion
{\displaystyle\oint}
%EndExpansion
dx\ \left\vert x\right\rangle \left\vert \varphi\left(  x\right)
\right\rangle
\]

\item[\fbox{\textbf{Step 3}}] Apply the Fourier transform to the left
register, i.e. apply $\mathcal{F}\otimes1$ to obtain
\[
\left\vert \psi_{3}\right\rangle =%
%TCIMACRO{\dsum \limits_{n\in\mathbb{Z}}}%
%BeginExpansion
{\displaystyle\sum\limits_{n\in\mathbb{Z}}}
%EndExpansion%
%TCIMACRO{\doint }%
%BeginExpansion
{\displaystyle\oint}
%EndExpansion
dx\ e^{-2\pi inx}\left\vert n\right\rangle \left\vert \varphi\left(  x\right)
\right\rangle =%
%TCIMACRO{\dsum \limits_{n\in\mathbb{Z}}}%
%BeginExpansion
{\displaystyle\sum\limits_{n\in\mathbb{Z}}}
%EndExpansion
\left\vert n\right\rangle
%TCIMACRO{\doint }%
%BeginExpansion
{\displaystyle\oint}
%EndExpansion
dx\ e^{-2\pi inx}\left\vert \varphi\left(  x\right)  \right\rangle
\in\mathcal{H}_{\mathbb{Z}}\otimes\mathcal{H}_{\mathbb{C}}%
\]
This can be shown to reduce to
\[
\left\vert \psi_{3}\right\rangle =%
%TCIMACRO{\dsum \limits_{\ell\in\mathbb{Z}}}%
%BeginExpansion
{\displaystyle\sum\limits_{\ell\in\mathbb{Z}}}
%EndExpansion
\left\vert \ell a\right\rangle \left\vert \Omega\left(  \ell a\right)
\right\rangle
\]
where
\[
\left\vert \Omega\left(  \ell a\right)  \right\rangle =%
%TCIMACRO{\dint \limits_{0}^{1/a}}%
%BeginExpansion
{\displaystyle\int\limits_{0}^{1/a}}
%EndExpansion
dx\ e^{-2\pi inx}\left\vert \varphi\left(  x\right)  \right\rangle \text{ .}%
\]

\end{itemize}

\bigskip

\begin{itemize}
\item[\fbox{\textbf{Step 4}}] Measure $\left\vert \psi_{3}\right\rangle =%
%TCIMACRO{\dsum \limits_{\ell\in\mathbb{Z}}}%
%BeginExpansion
{\displaystyle\sum\limits_{\ell\in\mathbb{Z}}}
%EndExpansion
\left\vert \ell a\right\rangle \left\vert \Omega\left(  \ell a\right)
\right\rangle $ with respect to the observable
\[
\mathcal{O}=%
%TCIMACRO{\dsum \limits_{n\in\mathbb{Z}}}%
%BeginExpansion
{\displaystyle\sum\limits_{n\in\mathbb{Z}}}
%EndExpansion
n\left\vert n\right\rangle \left\langle n\right\vert
\]
to produce a random eigenvalue $\ell a$.\ \bigskip

\item[\fbox{\textbf{Step 5}}] If the above steps are run twice to produce
eigenvalues $\ell_{1}a$ and $\ell_{2}a$, then the probability that the
integers $\ell_{1}$ and $\ell_{2}$ are relatively prime is $\zeta\left(
2\right)  ^{-1}\approx0.6079$, where $\zeta\left(  2\right)  $ denote the
Riemann zeta function at $k=2$. \ Hence, with high probability, the Euclidean
algorithm will produce the desired answer $a$.
\end{itemize}

\bigskip

\noindent\textbf{Remark.} \ \textit{Please note that, unlike Shor's algorithm
which uses the classical continued fraction algorithm in its last step to
determine the period, this algorithm uses in its last step only the much
faster classical Euclidean algorithm to find the period }$\frac{1}{a}%
$\textit{.}

\bigskip

\section{A curious algorithmic dual of Shor's algorithm}

\bigskip

Let us now construct a class of approximating QHS algorithms over finite
groups which asymptotically approach in the limit the QHS algorithm
\textsc{Alg}$_{\mathbb{R}/\mathbb{Z}}$. \ To do so, for each positive integer
$Q$, we approximate the infinite groups $\mathbb{Z}$ \ and $\mathbb{R}%
/\mathbb{Z}$ respectively by the finite groups
\[
\left\{
\begin{array}
[c]{ccc}%
\mathbb{Z} & \approx & \mathbb{Z}_{Q}=\left\{  k\in\mathbb{Z}%
\operatorname{mod}Q:0\leq k<Q\right\} \\
&  & \\
\mathbb{R}/\mathbb{Z} & \approx & \mathbb{Z}_{Q}=\left\{  \frac{r}%
{Q}:r=0,1,\ldots,Q-1\right\}
\end{array}
\right.
\]
and we approximate the map $\varphi:\mathbb{Z}\rightarrow\mathbb{C}$ by the
obvious map $\widetilde{\varphi}:\mathbb{Z}_{Q}\rightarrow\mathbb{C}$. \ The
resulting algorithm \textsc{Alg}$_{\mathbb{Z}_{Q}}$\textquotedblleft
lives\textquotedblright\ in the Hilbert space $\mathcal{H}_{\mathbb{Z}_{Q}%
}\otimes\mathcal{H}_{\mathbb{C}}$ and uses the approximating observable
\[
\mathcal{O}_{Q}=%
%TCIMACRO{\dsum \limits_{n=0}^{Q-1}}%
%BeginExpansion
{\displaystyle\sum\limits_{n=0}^{Q-1}}
%EndExpansion
n\left\vert n\right\rangle \left\langle n\right\vert
\]
The result for each $Q$ is a QHS algorithm $\textsc{Alg}_{\mathbb{Z}_{Q}}$
which is the algorithmic dual of Shor's algorithm. \ Because of the remark
found at the end of the previous section, it appears to run much faster. \ 

\bigskip

\section{A QHS Algorithm $\textsc{Alg}_{\text{\textsc{Paths}}}$ on the space
$\textsc{Paths}$}

\bigskip

The reader should be cautioned that the following algorithm is highly
speculative. This algorithm, in the spirit of Feynman, is based on functional
integrals whose existence is difficult to determine, let alone approximate. \ 

\bigskip

Let \textsc{Paths} be the space of all paths $x:\left[  0,1\right]
\rightarrow\mathbb{R}^{n}$ in real $n$-space $\mathbb{R}^{n}$ which are
$L^{2}$ with respect to the inner product
\[
x\cdot y=%
%TCIMACRO{\dint \limits_{0}^{1}}%
%BeginExpansion
{\displaystyle\int\limits_{0}^{1}}
%EndExpansion
ds\ x(s)y(s)
\]
We make \textsc{Paths} into a vector space over the reals $\mathbb{R}$ by
defining scalar addition and vector addition respectively as
\[
\left\{
\begin{array}
[c]{lll}%
\left(  \lambda x\right)  (s) & = & \lambda x(s)\\
&  & \\
\left(  x+y\right)  \left(  s\right)  & = & x(s)+y(s)
\end{array}
\right.
\]

We wish to solve the following problem:

\bigskip

\noindent\textbf{Problem.} \textit{Given a functional }$\varphi:$%
\textit{Paths}$\rightarrow R^{n}$\textit{\ and a hidden subspace }%
$V$\textit{\ of Paths such that }%
\[
\varphi(x+v)=\varphi\left(  x\right)  \ \forall v\in V\text{ ,}%
\]
\textit{create a QHS algorithm that finds the hidden subspace }$V$\textit{.}

\bigskip

Let $\mathcal{H}_{\text{\textsc{Paths}}}$ be the rigged Hilbert space with
orthonormal basis
\[
\left\{  \left\vert x\right\rangle :x\in\text{\textsc{Paths}}\right\}  \text{
}%
\]
where we have defined the bracket product as
\[
\left\langle x|y\right\rangle =\delta(x-y)
\]

Keeping in mind that the space Paths can be written as the disjoint union
\[
\text{\textsc{Paths}}=%
%TCIMACRO{\dbigcup \limits_{v\in V}}%
%BeginExpansion
{\displaystyle\bigcup\limits_{v\in V}}
%EndExpansion
\left(  v+V^{\bot}\right)  \text{ ,}%
\]
where $V^{\bot}$ denotes the orthogonal complement of $V$ in \textsc{Paths},
we proceed with the following QHS algorithm:

\bigskip

\bigskip

\begin{itemize}
\item[\fbox{\textbf{Step 0}}] Initialize
\[
\left\vert \psi_{0}\right\rangle =\left\vert 0\right\rangle \left\vert
0\right\rangle \in\mathcal{H}_{\text{$\textsc{Paths}$}}\otimes\mathcal{H}%
_{\mathbb{R}^{n}}%
\]

\item[\fbox{\textbf{Step 1}}] Apply the inverse Fourier transform to the left
register, i.e. apply $\mathcal{F}^{-1}\otimes1$ to obtain
\[
\left\vert \psi_{1}\right\rangle =%
%TCIMACRO{\dint \limits_{\text{\textsc{Paths}}}}%
%BeginExpansion
{\displaystyle\int\limits_{\text{\textsc{Paths}}}}
%EndExpansion
\mathcal{D}x\ e^{2\pi ix\cdot0}\left\vert x\right\rangle \left\vert
0\right\rangle =%
%TCIMACRO{\dint \limits_{\text{\textsc{Paths}}}}%
%BeginExpansion
{\displaystyle\int\limits_{\text{\textsc{Paths}}}}
%EndExpansion
\mathcal{D}x\ \left\vert x\right\rangle \left\vert 0\right\rangle
\]

\item[\fbox{\textbf{Step 2}}] Apply $U_{\varphi}:\left\vert x\right\rangle
\left\vert u\right\rangle \longmapsto\left\vert x\right\rangle \left\vert
u+\varphi\left(  x\right)  \right\rangle $ to obtain
\[
\left\vert \psi_{2}\right\rangle =%
%TCIMACRO{\dint \limits_{\text{\textsc{Paths}}}}%
%BeginExpansion
{\displaystyle\int\limits_{\text{\textsc{Paths}}}}
%EndExpansion
\mathcal{D}x\ \left\vert x\right\rangle \left\vert \varphi\left(  x\right)
\right\rangle
\]

\item[\fbox{\textbf{Step 3}}] Apply the Fourier transform to the left
register, i.e. apply $\mathcal{F}\otimes1$ to obtain
\[
\left\vert \psi_{3}\right\rangle =%
%TCIMACRO{\dint \limits_{\text{\textsc{Paths}}}}%
%BeginExpansion
{\displaystyle\int\limits_{\text{\textsc{Paths}}}}
%EndExpansion
\mathcal{D}y\
%TCIMACRO{\dint \limits_{\text{\textsc{Paths}}}}%
%BeginExpansion
{\displaystyle\int\limits_{\text{\textsc{Paths}}}}
%EndExpansion
\mathcal{D}x\ \ e^{-2\pi ix\cdot y}\left\vert y\right\rangle \left\vert
\varphi\left(  x\right)  \right\rangle =%
%TCIMACRO{\dint \limits_{\text{\textsc{Paths}}}}%
%BeginExpansion
{\displaystyle\int\limits_{\text{\textsc{Paths}}}}
%EndExpansion
\mathcal{D}y\ \left\vert y\right\rangle
%TCIMACRO{\dint \limits_{\text{\textsc{Paths}}}}%
%BeginExpansion
{\displaystyle\int\limits_{\text{\textsc{Paths}}}}
%EndExpansion
\mathcal{D}x\ \ e^{-2\pi ix\cdot y}\left\vert \varphi\left(  x\right)
\right\rangle
\]
Using the decomposition \textsc{Paths}$=%
%TCIMACRO{\dbigcup \limits_{v\in V}}%
%BeginExpansion
{\displaystyle\bigcup\limits_{v\in V}}
%EndExpansion
\left(  v+V^{\bot}\right)  $, we can formally show that this reduces to
\[
\left\vert \psi_{3}\right\rangle =%
%TCIMACRO{\dint \limits_{V^{\bot}}}%
%BeginExpansion
{\displaystyle\int\limits_{V^{\bot}}}
%EndExpansion
\mathcal{D}u\ \left\vert y\right\rangle \left\vert \Omega\left(  u\right)
\right\rangle \text{ ,}%
\]
where
\[
\left\vert \Omega\left(  u\right)  \right\rangle =%
%TCIMACRO{\dint \limits_{V^{\bot}}}%
%BeginExpansion
{\displaystyle\int\limits_{V^{\bot}}}
%EndExpansion
\mathcal{D}x\ e^{-2\pi ix\cdot u}\left\vert \varphi\left(  x\right)
\right\rangle \text{ .}%
\]

\end{itemize}

\bigskip

\begin{itemize}
\item[\fbox{\textbf{Step 4}}] Measure $\left\vert \psi_{3}\right\rangle =%
%TCIMACRO{\dint \limits_{V^{\bot}}}%
%BeginExpansion
{\displaystyle\int\limits_{V^{\bot}}}
%EndExpansion
\mathcal{D}u\ \left\vert y\right\rangle \left\vert \Omega\left(  u\right)
\right\rangle $ with respect to the observable
\[
\mathcal{O}=%
%TCIMACRO{\dint \limits_{\text{\textsc{Paths}}}}%
%BeginExpansion
{\displaystyle\int\limits_{\text{\textsc{Paths}}}}
%EndExpansion
\mathcal{D}w\ w\left\vert w\right\rangle \left\langle w\right\vert
\]
to produce a random element of $V^{\bot}$.
\end{itemize}

\bigskip

\section{More speculations and questions}

\bigskip

Can the above highly speculative algorithm be modified in such a way to create
a QHS algorithm for the Jones polynomial? In other words, can it be modified
by replacing the space \textsc{Paths} with the space $\mathcal{A}$ of gauge
connections, and by making suitable modifications of the functional integral
$\ $%
\[
\widehat{\psi}\left(  K\right)  =%
%TCIMACRO{\dint \nolimits_{\mathcal{A}}}%
%BeginExpansion
{\displaystyle\int\nolimits_{\mathcal{A}}}
%EndExpansion
\mathcal{D}A\psi(A)\mathcal{W}_{K}(A)
\]
where $\mathcal{W}_{K}(A)$ denotes the Wilson loop
\[
\mathcal{W}_{K}(A)=tr\left(  P\exp\left(
%TCIMACRO{\doint _{K}}%
%BeginExpansion
{\displaystyle\oint_{K}}
%EndExpansion
A\right)  \right)
\]
\bigskip

The functional integral $\widehat{\psi}\left(  K\right)  $ transforms the
function $\psi\left(  A\right)  $ of gauge connections to a function of closed
curves in three dimensional space. Witten\cite{Witten1} showed that, if
$\psi\left(  A\right)  $ is chosen correctly (an exponentiated integral of the
Chern-Simons form), then, up to framing corrections, $\widehat{\psi}\left(
K\right)  $ is a knot and link invariant. This invariant reproduces the
original Jones polynomial for appropriate choice of the gauge group, and many
other invariants for other such choices. By continuing our exploration of
quantum algorithms as in the last section, we hope to give a quantum algorithm
for these topological invariants, thereby forging a new connection between
quantum computing and topological quantum field theory.

\bigskip

\section{Acknowledgements}

\bigskip

\bigskip

This effort is partially supported by the Defense Advanced Research Projects
Agency (DARPA) and Air Force Research Laboratory, Air Force Materiel Command,
USAF, under agreement number F30602-01-2-0522, the National Institute for
Standards and Technology (NIST), and by L-O-O-P Fund Grant BECA2002. The U.S.
Government is authorized to reproduce and distribute reprints for Government
purposes notwithstanding any copyright annotations thereon. The views and
conclusions contained herein are those of the authors and should not be
interpreted as necessarily representing the official policies or endorsements,
either expressed or implied, of the Defense Advanced Research Projects Agency,
the Air Force Research Laboratory, or the U.S. Government. The first author
gratefully acknowledges the hospitality of the Mathematical Science Research
Institute, Berkeley, California, where some of this work was completed, and
would also like to thank Alexei Kitaev for some helpful conversations.

\end{document}